\documentclass[10pt,twocolumn,twoside]{article}

\usepackage[utf8]{inputenc}   % <<<<< Linux
\usepackage[english]{babel} % <<<<< English
\usepackage[T1]{fontenc}

\usepackage{geometry}	
\usepackage{setspace}

\usepackage{graphicx}

\usepackage{amsmath}  % AMS mathematical facilities for LaTeX.
\usepackage{amsthm}   % Typesetting theorems (AMS style).
\usepackage{amsfonts} % 
\usepackage{amssymb}

 \usepackage{wrapfig}

\usepackage{subfigure}

\usepackage{subfigmat}

 \usepackage{url}

\usepackage{varioref}

\usepackage{dcolumn}
\newcolumntype{d}{D{.}{.}{-1}} % column aligned by the point separator '.'
\newcolumntype{e}{D{E}{E}{-1}} % column aligned by the exponent 'E'

 \usepackage{verbatim}

\usepackage{rotating}

\usepackage[pdftex]{hyperref} % enhance documents that are to be
\hypersetup{colorlinks,       % color text of links and anchors,
            linkcolor=black,  % color for normal internal links
            anchorcolor=black,% color for anchor text
            citecolor=black,  % color for bibliographical citations
            filecolor=black,  % color for URLs which open local files
            menucolor=black,  % color for Acrobat menu items
            pagecolor=black,  % color for links to other pages
            urlcolor=blue,   % color for linked URLs
	          bookmarks=true,         % create PDF bookmarks
	          bookmarksopen=false,    % don't expand bookmarks
	          bookmarksnumbered=true, % number bookmarks
	         }

\usepackage[figure,table]{hypcap}
\usepackage{float}

\usepackage[numbers,sort&compress]{natbib} % <<<<< References in numbered list [1],[2],...
\usepackage{notoccite}

\usepackage{multirow}

\usepackage{booktabs}
\usepackage{pdfpages}

\usepackage{csquotes}
\usepackage{nameref}  %for named cross-references

\usepackage{tikz}
\usetikzlibrary{positioning}
\usepackage{ragged2e}  %justify text

\usepackage{enumitem}

\usepackage{diagbox}

\usepackage{appendix}

\usepackage{nicefrac, xfrac}

\usepackage{makecell}  % used in table

\usepackage{colortbl}

\usepackage[bf,small]{caption}

\DeclareCaptionType{equ}[][]
\usepackage[export]{adjustbox}

\usepackage{hhline}

\usepackage{fancyhdr}

\usepackage{array}

\usepackage{authblk}

\usepackage{hyperref}

\usepackage{multicol}

 \usepackage{pdfpages}

\begin{document}

\def\methods{\textbf{\sffamily Material and Methods}}
\def\si{\textbf{\sffamily Supporting Information}}
\newcommand{\christian}[1]{\textcolor{blue}{\textbf{CH}: #1}}

\title{Evolution of noisy learning in games}

% Use letters for affiliations, numbers to show equal authorship (if applicable) and to indicate the corresponding author
\author[a,b,1]{Marta C. Couto}
\author[a]{Fernando P. Santos}
\author[b,c]{Christian Hilbe}

\affil[a]{\footnotesize{Informatics Institute, University of Amsterdam, Amsterdam 1098XH, The Netherlands}}
\affil[b]{Max Planck Research Group on the
Dynamics of Social Behavior, Max Planck Institute for
Evolutionary Biology, 24306 Pl\"{o}n, Germany}
\affil[c]{Interdisciplinary Transformation University IT:U, 4040 Linz, Austria}
\affil[1]{To whom correspondence should be addressed. E-mail: m.gomesdacunhacouto@uva.nl, mccouto@evolbio.mpg.de}

%\maketitle

\twocolumn[
\begin{@twocolumnfalse}
	\maketitle
	\begin{abstract}
People make strategic decisions many times a day -- during negotiations, when coordinating actions with others, or when choosing partners for cooperation.
The resulting dynamics can be studied with learning theory and evolutionary game theory. 
These frameworks explore how people adapt their decisions over time, in light of how effective their strategies have been. 
The outcomes of such learning processes depend on how sensitive individuals are to the performance of their strategies. 
When they are more sensitive, they systematically favor strategies they deem more successful.
When they are less sensitive, their learning process is noisier and more erratic.
Traditionally, most models treat this sensitivity as a fixed parameter -- like the `selection strength' parameter in evolutionary models. 
Instead, we study how strategies and sensitivities co-evolve.  
We find that the co-evolutionary endpoints depend on both the type of strategic interaction and the learning rule employed. 
In prisoner’s dilemmas, we often observe sensitivities to increase indefinitely. 
But in snowdrift and stag-hunt games, sensitivities often converge to a finite value, or we observe evolutionary branching altogether. 
These results shed light on how evolution might shape learning mechanisms for social behavior. 
They suggest that noisy learning does not need to be a by-product of cognitive constraints. 
Instead, it can serve as a means to gain strategic advantages.\\
%\vspace{0.3cm}
\vspace{1cm}
	\end{abstract}
\end{@twocolumnfalse}
]

%%%%%%%%%%%%%%
%%  INTRODUCTION  %%
%%%%%%%%%%%%%%

\section*{Introduction}

{P}eople constantly make strategic decisions. 
These decisions range from inconsequential choices (e.g., whether to do a small favor) to major commitments (e.g., choosing a business partner).
The process by which individuals arrive at their decisions can be analyzed with learning theory~\cite{macy:PNAS:2002,Galla:PNAS:2013,Fudenberg:book:1998b} or evolutionary game theory~\cite{Nowak:Science:2004,Broom:book:2013,Traulsen:PTRSB:2023, McNamara:Nature:1999, Mcnamara:JRSI:2013, Leimar:SciRep:2019}. 
To this end, individuals are regarded as players who interact in games. 
They can choose among different strategies and win a payoff that depends both on their own and their co-players' decisions. 
Importantly, they do not need to behave optimally from the outset. 
Rather, individuals are assumed to engage in the same interaction (or sufficiently similar ones) many times. 
Over the course of these interactions, they can adapt their behavior to yield better payoffs. 

\begin{figure*}[t]
	\centering
	\includegraphics[width=\textwidth]{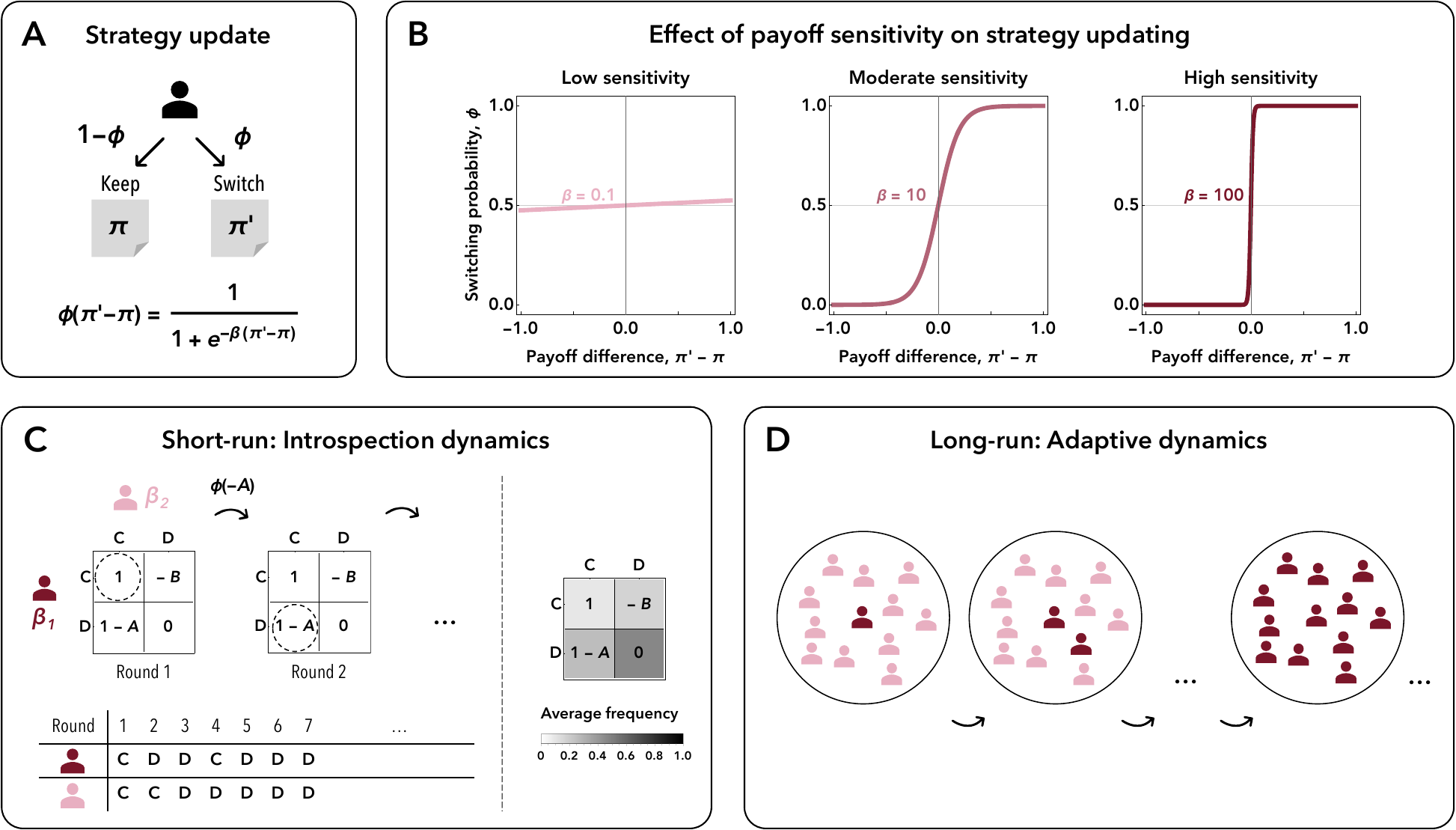}\\[-0.2cm]
	\caption{
		\textbf{An overview of the model.} 
		\textbf{A,} In stochastic models of evolutionary game theory, individuals continually get a chance to revise their strategy.
		We model this revision process with introspection dynamics~\cite{Couto:NJP:2022, Couto:DGAA:2023,Hauser:Nature:2019,McAvoy:PNASnexus:2022,Wang:PTRSB:2023,huebner:PNAS:2024,Schmid:PlosCB:2022,Ramirez:SciRep:2023}. 
		Here, a player compares its current strategy to a randomly chosen alternative. 
		If $\pi$ is the payoff of the current strategy, and $\pi'$ the payoff of the alternative, the player's switching probability is given by a Fermi function $\phi(\pi' \!-\! \pi)$. 
		This implies players are more likely to switch if the alternative strategy is more beneficial. 
		\textbf{B,} The exact shape of the switching probability depends on a parameter $\beta$. 
		This parameter is often referred to as the strength of selection. Because here we interpret $\beta$ as an individual (and evolvable) trait, we refer to it as the player's \textit{payoff sensitivity}. 
		For low~$\beta$, the player's learning process is more noisy. 
		Here, strategy changes are mostly driven by chance. 
		As $\beta$ becomes larger, updating decisions become more deterministic. Here, individuals increasingly favor those alternative strategies with high payoffs. 
		We are interested in the evolution of this parameter $\beta$. Our process unfolds on two timescales.
		\textbf{C,} In the short run, each player's payoff sensitivity $\beta$ is fixed. 
		Given their sensitivity, they choose between different strategies in a stage game. 
		The depicted example shows a round in which player 1 switches from strategy C (cooperation) to strategy D (defection). 
		We iterate this learning process for many rounds. 
		Based on these iterations, we compute the average frequency of each possible game outcome (illustrated as a black-and-white gradient). 
		This allows us to compute the players' expected payoff as a function of their $\beta$ values. 	
		\textbf{D,} In the long-run, we let the players' payoff sensitivity~$\beta$ evolve. 
		We model this long-run process with adaptive dynamics~\cite{Geritz:PRL:1997, geritz:EER:1998, Hofbauer:book:1998,Brannstrom:Games:2013}.}
	\label{Fig1}
\end{figure*}

The outcome of this adaptation process depends on how sensitive players are to their strategies' performance. 
Most models capture this sensitivity by a parameter. 
For example, in stochastic learning models of evolutionary game theory, outcomes crucially depend on the \textit{selection strength} or \textit{intensity of selection},~$\beta$. 
This parameter~$\beta$ determines to which extent individuals switch their strategies because of payoff differences~(\textbf{Fig.~\ref{Fig1}A}). 
When this parameter is small, players are largely insensitive to their payoffs. 
They mostly switch from one strategy to another by chance~(\textbf{Fig.~\ref{Fig1}B}, left).
The respective limit of weak selection ($\beta\!\rightarrow\!0$) has become a major paradigm in evolutionary game theory, partly because it simplifies many calculations~\citep{Nowak:Nature:2004,Tarnita:JTB:2009,McAvoy:JMB:2015,Allen:Nature:2017,Su:PNAS:2022}.
In this limit, learning takes the form of an (almost unbiased) random walk on the space of all strategies.
In contrast, as selection strength becomes large, players only switch strategies if the new strategy is superior~(\textbf{Fig.~\ref{Fig1}B}, right). 
The respective limit of strong selection~($\beta\!\rightarrow\!\infty$) has become another major paradigm to study evolutionary games~\citep{Hauert:Science:2007,Sigmund:Nature:2010,Garcia:PLoSOne:2012}. 
Respective results are closely aligned with key concepts in classical game theory.
For example, in this limit, strict Nash equilibria correspond to absorbing states of the learning process~\citep{Fudenberg:TPB:2006,Fudenberg:JET:2008}. 
In the following, we refer to $\beta$ as a player's (\textit{payoff}) \textit{sensitivity}.
A small $\beta$ translates into a more noisy learning process.
A large $\beta$ makes individuals update their strategies more deterministically. 
Related parameters also exist in other learning models. 
For example, in experience-weighted attraction learning~\cite{Camerer:Econometrica:1999,Dridi:TheoPopBiol:2014,Pangallo:GEB:2022}, outcomes critically depend on a player's {\it sensitivity to attractions}.

%[raising the research questions: (1) asymmetric selection strength]
Except for a handful of studies discussed further below~\cite{Szabo:EPL:2009, Szolnoki:PRE:2009c}, most models interpret a player's (payoff) sensitivity as a fixed quantity, often assumed to be the same for everyone.  
Here, we relax this assumption. 
We address two related questions.
First, we are interested in learning processes among individuals with fixed but different sensitivities. 
We refer to individuals with larger~$\beta$ as more sensitive learners, and those with lower~$\beta$ as less sensitive learners.  
We ask to what extent more sensitive learners are able to systematically outperform their opponents. 

%[raising the research questions: (2) evolution of selection strength]
Second, we explore the dynamics when a player's sensitivity is itself an evolving trait. 
To this end, we consider a framework akin to the indirect evolutionary approach used to study the evolution of preference~\cite{Guth:RS:1998,Huck:GEB:1996, Akcay:PNAS:2009, Alger:PTRSB:2023, Wang:PNAS:2024}. 
The dynamics unfolds on two time scales. 
In the short-run, sensitivities are fixed. 
Players interact in a given game with others, adopt new strategies based on their sensitivity, and as a result, earn a certain average payoff~(\textbf{Fig.~\ref{Fig1}C}). 
In the long-run, we allow their sensitivity to evolve, based on the average payoffs they yield. 
To the extent that a player's sensitivity is an inheritable trait, this assumption may reflect that more effective learners have more offspring~(\textbf{Fig.~\ref{Fig1}D}). 
We explore this long-run dynamics analytically and with simulations, based on the adaptive dynamics formalism~\cite{Geritz:PRL:1997, geritz:EER:1998, Hofbauer:book:1998,Brannstrom:Games:2013}. 

To get an intuitive understanding, we focus on the simplest class of games: symmetric games among two players with two strategies. 
These games include the prisoner's dilemma, the stag-hunt game, and the snowdrift game. 
However, we also show that the respective intuitions carry over to more general settings, including multiplayer and repeated games. 
Similarly, throughout most of the main text, we study a baseline model based on one particular learning rule, called {\it introspection dynamics}~\cite{Couto:NJP:2022, Couto:DGAA:2023,Hauser:Nature:2019,McAvoy:PNASnexus:2022,Wang:PTRSB:2023,huebner:PNAS:2024,Schmid:PlosCB:2022,Ramirez:SciRep:2023}. 
This dynamics is particularly convenient to work with because it allows for a fully analytical treatment. 
However, further below (and in more detail in the \si), we discuss to which extent the respective results generalize to other learning rules. 

%[results summary]
Once a player's sensitivity is an evolvable trait, one might expect an evolutionary dynamics towards ever-increasing sensitivities. 
After all, more sensitive learners are better at identifying strategies with high payoffs.
Hence, such learners might have a fitness advantage.
Remarkably, however, we do not find such a runaway dynamics for all games. 
Instead, in many snowdrift games, the players' sensitivities converge to a finite value. 
Moreover, in some stag-hunt games, we observe evolutionary branching. 
Here, an initially monomorphic population splits into two or more subpopulations, each with a different sensitivity. 
Interestingly, we obtain these results without assuming that more sensitive learners exert more cognitive effort and hence pay some complexity cost.
Instead, we obtain these results because a noisy learning process can turn out to be a strategic advantage. 
In this way, our study contributes to a growing literature exploring how strategic considerations can explain behaviors that are seemingly at odds with sensible decision-making~\cite{Guth:RS:1998,Huck:GEB:1996, Akcay:PNAS:2009, Alger:PTRSB:2023, Wang:PNAS:2024}.

%%%%%%%%%%%%%%%%%
%% MODEL AND RESULTS %%
%%%%%%%%%%%%%%%%%

\section*{Model and results}
Our model assumes a separation of timescales. 
In the short run, players have fixed but possibly different payoff sensitivities. 
They repeatedly engage in a given one-shot game, and they learn to adopt more profitable strategies over time. 
We model this short-run learning process with introspection dynamics~\cite{Couto:NJP:2022, Couto:DGAA:2023,Hauser:Nature:2019,McAvoy:PNASnexus:2022,Wang:PTRSB:2023,huebner:PNAS:2024,Schmid:PlosCB:2022,Ramirez:SciRep:2023}. 
In the long run, the players' payoff sensitivities are allowed to evolve. 
We describe this long-run process with adaptive dynamics~\cite{Geritz:PRL:1997, geritz:EER:1998, Hofbauer:book:1998,Brannstrom:Games:2013}, a classical tool to study the evolution of continuous traits.
In the following, we describe the two processes in turn.

%% SHORT-RUN DYNAMICS %%

\subsection*{A description of the short-run dynamics}
\label{Ch3_Model_IntDyn}
To explore learning among players with different payoff sensitivities, we consider simple one-shot, $2\times2$ games.
That is, we consider two players who can choose among two possible strategies, $\mathbf{C}$ and $\mathbf{D}$ (we think of the two strategies as cooperation and defection, but this interpretation is irrelevant for our results). 
The choices of the two players determine their payoffs. 
Payoffs are assumed to be symmetric, and given by the (normalized) payoff matrix
\begin{equation}
	\label{payoffmatrix}
	\begin{array}{c|cc}
		&\mathbf{\;C\;} &\mathbf{\;D\;}\\
		\hline
		\mathbf{\;C\;} &1 &-B \\
		\mathbf{\;D\;} &1-A &0 \\
	\end{array}
\end{equation}
We refer to this matrix as the stage game.
As usual, player~1 chooses a row of this matrix, player~2 chooses a column, and the entries represent the payoffs of player~1. 
Depending on the signs of $A$ and $B$, we recover four major game classes: 
prisoner's dilemma~($A\!<\!0$,  $B\!>\!0$), stag-hunt game ($A\!>\!0$, $B\!>\!0$), snowdrift game ($A\!<\!0$, $B\!<\!0$), and harmony game~($A\!>\!0$, $B\!<\!0)$. 

% [Description of learning process]
To describe how players adopt more profitable strategies over time, we use introspection dynamics \cite{Couto:NJP:2022, Couto:DGAA:2023}.
This process assumes that the two individuals interact in the same stage game over many time steps.  
Each time step, a randomly chosen player is given a chance to revise its strategy. 
That player then compares its current payoff~$\pi$ with the payoff $\pi'$ the player could have obtained by choosing the other strategy (keeping the co-player's strategy fixed). 
As illustrated in \textbf{Fig.~\ref{Fig1}A}, the player switches to the other strategy with a probability that depends on the payoff difference~\cite{Blume:GEB:1993, szabo:PRE:1998, Traulsen:PRE:2006b},
\begin{equation}
	\label{comparisonrule}
	\phi=\frac{1}{1+{\rm e}^{-\beta_i\, (\pi' - \pi)}}.
\end{equation}
\noindent The parameter $\beta_i\!\geq\!0$ represents player $i$'s {payoff sensitivity}. 
For small $\beta_i$, this function approaches $1/2$ irrespective of the payoff difference~(\textbf{Fig.~\ref{Fig1}B}~left). 
In this limit, players decide whether to switch strategies based on a coin toss. 
For large $\beta_i$, the function approaches a step function~(\textbf{Fig.~\ref{Fig1}B} right). 
Here, the player always switches if the alternative strategy yields the better payoff. 
We iterate this basic updating process for many time steps. 
As a result, players repeatedly get a chance to switch their strategies~~(\textbf{Fig.~\ref{Fig1}C}, bottom). 
Based on this process, we record how frequently we observe each of the four possible game outcomes over time. 
In the right panel of \textbf{Fig.~\ref{Fig1}C}, we illustrate these overall frequencies using a black-and-white gradient. 
A darker shading indicates game outcomes that are observed more often. 

% [Explicit formulas]
Introspection dynamics is particularly convenient because the expected frequencies of the four game outcomes can be computed analytically for any values of $\beta$ -- contrary to other stochastic dynamics such as pairwise imitation \cite{Traulsen:PRE:2006b} or experienced-weighted attraction learning models \cite{Camerer:Econometrica:1999}.
To compute the expected frequencies of each game outcome, we represent the learning dynamics as a Markov chain (see \si{} for details). 
The Markov chain has four states, corresponding to the four possible outcomes in each time step, \textbf{CC}, \textbf{CD}, \textbf{DC}, \textbf{DD} (the first letter refers to the first player's strategy, and the second letter to the second's). 
For finite $\beta_1, \beta_2$, this Markov chain has a unique stationary distribution $\mathbf{u}(A, B, \beta_1, \beta_2)~:=~(u_{\textbf{CC}}, u_{\textbf{CD}}, u_{\textbf{DC}}, u_{\textbf{DD}})$. 
This distribution captures how often we observe each game outcome on average, over the course of many rounds. 
Based on the stationary distribution, it is straightforward to compute player 1's average payoff. 
To this end, we multiply the four frequencies $u_{ij}$ with the respective stage game payoffs, 
\begin{equation} 
	\Pi_{\beta_2}(\beta_1) = 1\!\cdot\!u_\textbf{CC} -B\!\cdot\! u_\textbf{CD} +(1\!-\!A)\!\cdot\! u_\textbf{DC}+0\!\cdot\! u_\textbf{DD}.
	\label{averagepayoffs}
\end{equation}

% [The symmetric case]
\noindent
The respective results become particularly simple when both players have the same payoff sensitivity,~$\beta_1\!=\!\beta_2$. 
In that case, one can show that the stationary distribution simplifies to
\begin{equation} 
	\mathbf{u}(A, B, \beta, \beta)= \frac{1}{2 + e^{A \beta} + e^{B \beta}}(e^{A \beta},1,1,e^{B \beta}). 
	\label{stationarydistributionsymmetric}
\end{equation}
For weak selection  ($\beta\!\rightarrow\!0$), this formula implies that all game outcomes are observed equally often, $\mathbf{u}=(\frac{1}{4}, \frac{1}{4}, \frac{1}{4}, \frac{1}{4})$. 
For strong selection ($\beta\!\rightarrow\!\infty$), the stationary distribution depends on the game parameters $A$ and $B$. 
For example, for the prisoner's dilemma,  \eqref{stationarydistributionsymmetric} implies that the two players defect in almost all rounds ($u_{\textbf{DD}} \!=\! 1$); 
in the snowdrift game, they anti-coordinate ($u_\textbf{CD}\!=\!u_{\textbf{DC}}\!=\!\frac{1}{2}$);
and in the stag-hunt game they choose the strategy that is risk-dominant \citep[$u_{\textbf{DD}}\!=\! 1$ if $A\!<\!B$, $u_{\textbf{CC}}\!=\!1$ if $A>B$, see Ref.][]{Couto:NJP:2022}.

\begin{figure*}[t]
	\centering
	\includegraphics[width = 0.95\textwidth]{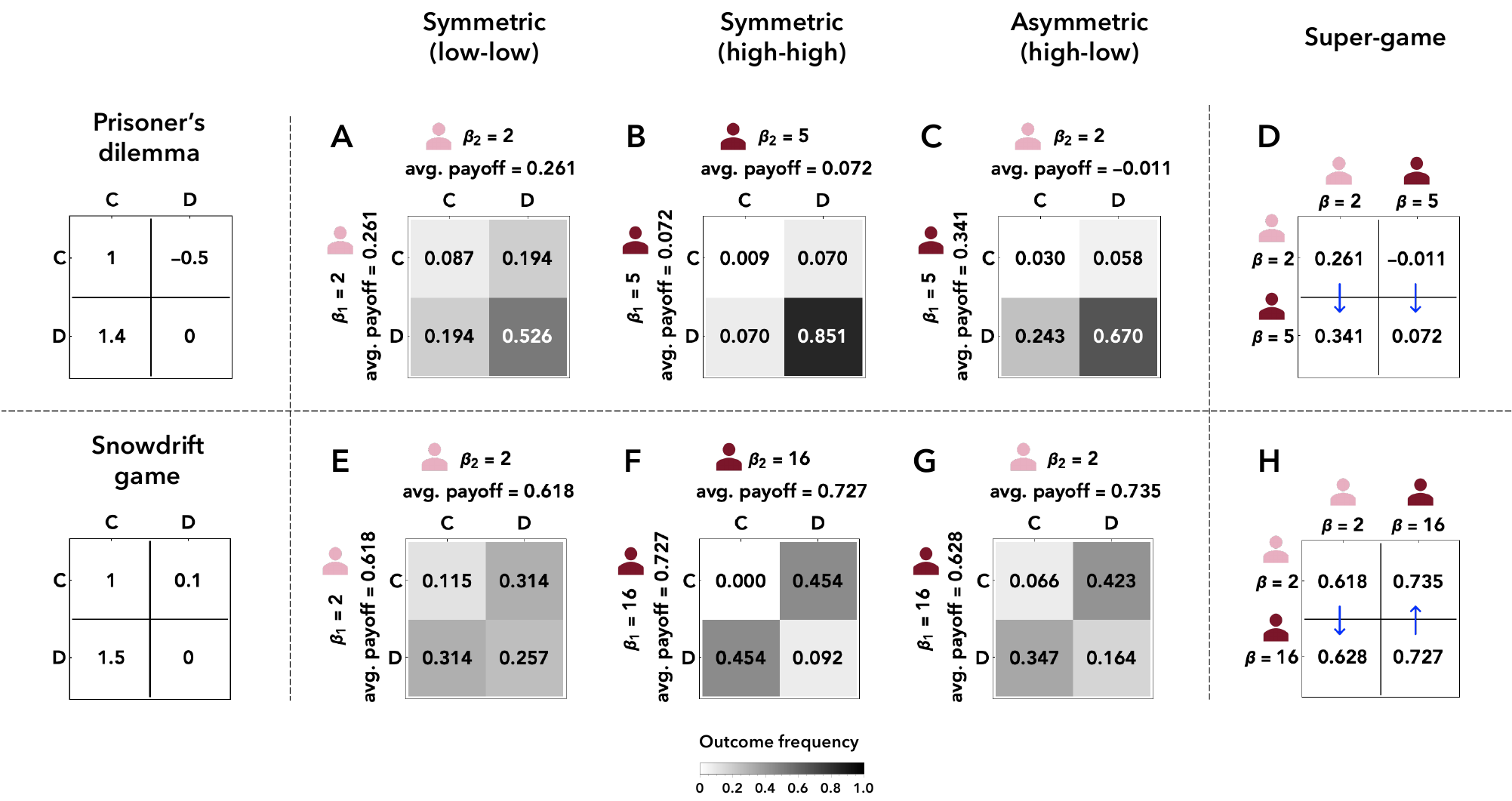}
	\caption{
		\textbf{Introspection dynamics among players with different payoff sensitivities.}
		To illustrate the impact of payoff sensitivity on the learning dynamics, we consider two stage games, a prisoner's dilemma (\textbf{A--C}) and a snowdrift game (\textbf{D--F}). In each case, we assume payoff sensitivities may take two possible values.
		They can either be comparably low (bright avatar) or high (dark avatar). 
		For all possible combinations of payoff sensitivities, we depict how often the respective players would obtain one of the four possible payoffs of the respective stage game. 
		Based on this stationary distribution, we compute the players' expected payoffs. 
		For the prisoner's dilemma, we observe that the higher payoff sensitivity dominates the lower payoff sensitivity (indicated by blue arrows in panel \textbf{D}). 
		In contrast, for the snowdrift game, no payoff sensitivity value is dominant. Instead, each player prefers to have the opposite payoff sensitivity value of their opponent.
	}
	\label{Fig2}
\end{figure*}

% [The asymmetric case: prisoner's dilemma]
Using the same formalism, we can also analyze players with different payoff sensitivities. 
In the \si, we derive a general expression of the stationary distribution $\mathbf{u}(A, B, \beta_1, \beta_2)$ as a function of $\beta_1$ and $\beta_2$, which is now more complex. 
In the following, we use that expression to explore whether a more sensitive learner (one with a larger $\beta$) would gain a long-term payoff advantage in various stage games.

In \textbf{Fig.~\ref{Fig2}A--D}, we consider a prisoner's dilemma. 
We assume the two players can either have a small or a comparably large payoff sensitivity, $\beta\!\in\!\{2,5\}$. 
For a symmetric situation in which both players use the same $\beta$, we find that players learn to defect more often when $\beta$ is large~(\textbf{Fig.~\ref{Fig2}A,B}).
This result is intuitive: the larger the payoff sensitivity, the better players are at identifying the individually optimal strategy of defecting. 
Once players are asymmetric, we find that it is the more sensitive learner who defects more often~(\textbf{Fig.~\ref{Fig2}C}).  
As a result, this player ends up with a better average payoff.

% [The asymmetric case: snowdrift game]
In \textbf{Fig.~\ref{Fig2}E--H}, we consider a snowdrift game. 
For illustration, the possible sensitivities are now $\beta\!\in\!\{2,16\}$.
In the symmetric case among equally sensitive players, a large $\beta$ again makes it more likely that players implement a Nash equilibrium (here, \textbf{CD} or \textbf{DC}). 
In the depicted example, they coordinate on a Nash equilibrium in $63\%$ of all rounds when both have $\beta\!=\!2$, but in $91\%$ of all rounds when both have $\beta\!=\!16$. 
Remarkably, however, once players differ in their sensitivities, it is the {\it less} sensitive (and hence more erratic) learner who obtains the larger payoff ~(\textbf{Fig.~\ref{Fig2}G}).
The intuition is the following.  
When in their less preferred equilibrium \textbf{DC}, the less sensitive player is still relatively likely to deviate. 
For the given payoff values, the respective switching probability according to Eq.~[\ref{comparisonrule}] is $1/(1+{\rm e}^{2\times0.1})=0.45$.
This probability is smaller than $1/2$ (because the switch reduces the player's payoff from 0.1 to 0), but it remains substantial. 
After this switch towards \textbf{DD}, the more sensitive player is tempted to react by playing \textbf{C}. 
As a result, the two players transitioned from one Nash equilibrium (\textbf{DC}, which favors the more sensitive player) to another (\textbf{CD}, which favors the less sensitive player). 
Interestingly, the converse transition from \textbf{CD} to \textbf{DC} is less likely because, starting from \textbf{CD}, the less sensitive player now has more to lose by deviating (from 1.5 to 1.0).
%whereas the more sensitive player is unwilling to even accept a small and possibly temporary payoff loss. 

% [Summary and outlook]
Based on the above results, we can construct a `super-game' matrix~(\textbf{Fig.~\ref{Fig2}D,H}). 
This matrix represents the interaction of two learners who may have different payoff sensitivities. 
The entries of the matrix depict the learners' average payoffs over the course of the learning process (as derived in the previous panels).  
For the prisoner's dilemma, the super-game indicates that each learner always prefers to have a higher payoff sensitivity (indicated by blue arrows).
In the snowdrift game example, the super-game takes the form of an anti-coordination game: players prefer their payoff sensitivity to be high when their opponent's is low, and vice versa. 

Overall, these two examples provide a proof of principle. 
They illustrate the non-trivial effects of asymmetric payoff sensitivities. 
In general, the structure of the super-game depends on both the chosen $\beta$ values and the exact game parameters A and B. In particular, for the snowdrift game depicted in \textbf{Fig.~\ref{Fig2}}, one can also construct super-games in which the lower sensitivity dominates the larger one (or the other way round). Thus, these observations call for a more systematic analysis, considering more games and a larger set of possible sensitivities. We provide such an analysis in \textbf{Fig.~S1} of the \si{}.
In addition, the above results raise the question of whether the players' sensitivities would necessarily move towards ever-increasing values if it were an evolving trait. 
We explore that issue in the following.

%% LONG-RUN DYNAMICS %%

\begin{figure*}[t!]
	\centering
	\includegraphics[width =  0.95\textwidth]{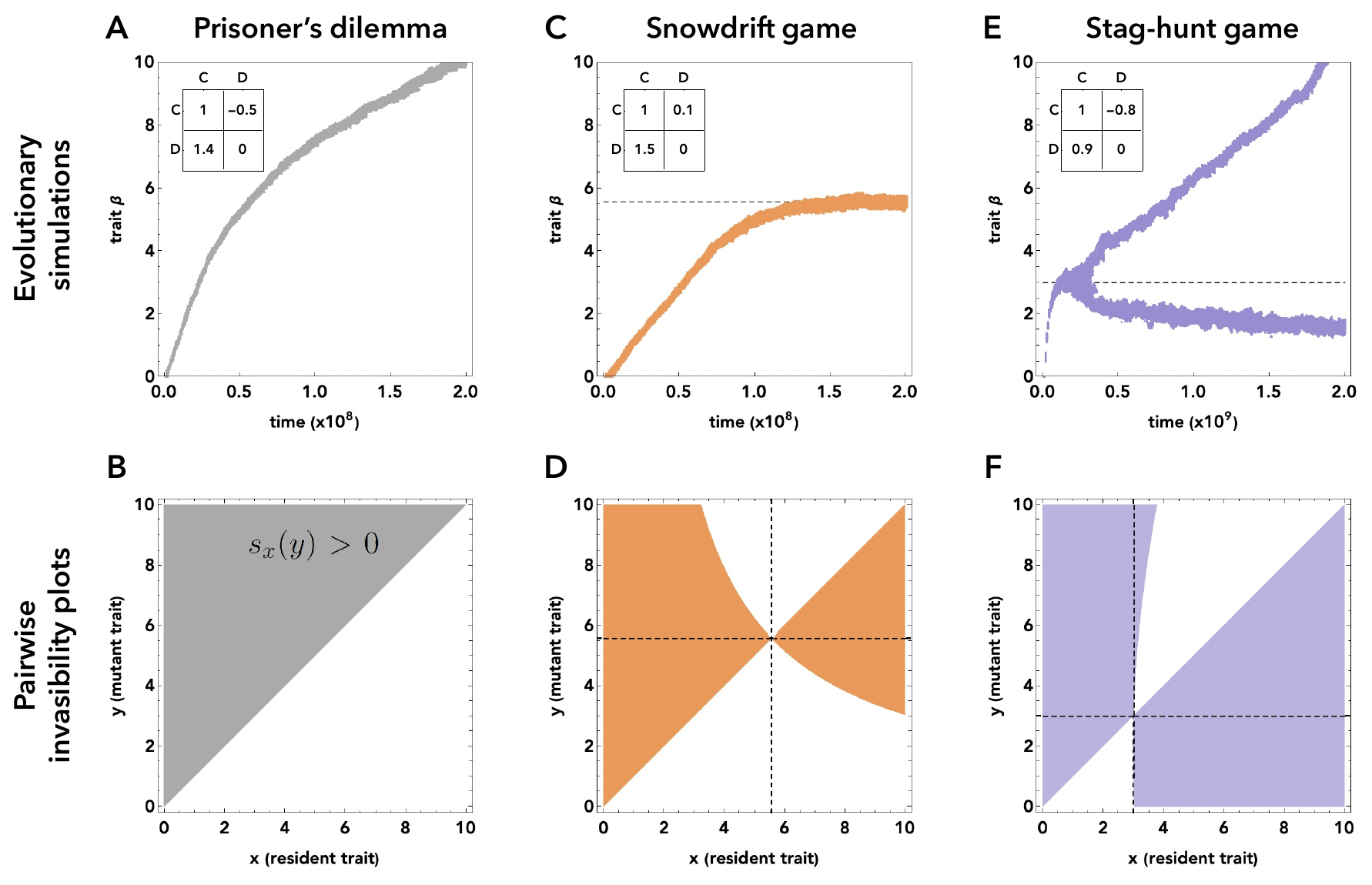}
	\caption{
		\textbf{Adaptive dynamics of payoff sensitivity for three different social dilemmas.}
		We explore the adaptive dynamics for three different social dilemmas: the prisoner's dilemma, the snowdrift game, and the stag-hunt game. 
		For each case, we depict a realization of an individual-based simulation (upper panels) and a pairwise invasibility plot (lower panels). 
		Each simulation is initialized at a monomorphic population with a payoff sensitivity of $\beta\!=\!0$. 
		Over time, mutations introduce variation in the players' payoff sensitivity. 
		Individuals obtain payoffs by randomly interacting with other population members and learning strategies through introspection dynamics. 
		They reproduce according to their payoff. 
		For details, see \methods. 
		Pairwise invasibility plots illustrate the dynamics of monomorphic resident populations.  
		They display (in color) which rare mutant traits ($y$-axis) have a positive invasion fitness $s_x(y)$, given the current resident ($x$-axis). 
		Colored regions above the diagonal indicate that mutants with a higher trait value than the resident are favored to invade. 
		Colored regions below the diagonal indicate evolution towards smaller trait values. 
		Dashed lines indicate the position of singular trait values. 
		\textbf{A,B,} For the depicted prisoner's dilemma, we observe a dynamics of ever-increasing payoff sensitivities. 
		\textbf{C,D,} In the snowdrift game, the evolving $\beta$ values converge to a finite value. 
		\textbf{E,F,} In the stag-hunt game, we observe evolutionary branching. After this occurs, there are two subpopulations. 
		Members of the first subpopulation have low $\beta$ values, and hence they choose strategies mostly at random. 
		Members of the second subpopulation exhibit high $\beta$ values; they tend to best respond to their respective opponent.}
	\label{Fig3}
\end{figure*}

\subsection*{A description of the long-run dynamics}
To study how the players' payoff sensitivities $\beta$ might change over time, we use adaptive dynamics~\cite{Geritz:PRL:1997, geritz:EER:1998, Hofbauer:book:1998}. 
That is, we consider a large and well-mixed population of players~(\textbf{Fig.~\ref{Fig1}D}).
Each player randomly meets other population members to interact in a series of pairwise games. 
As a result, they obtain a payoff as described in the previous section on introspection dynamics. 
We define a player's fitness to be its average payoff across all its interaction partners. 
Players with larger fitness are more likely to reproduce. 
The offspring of a reproducing player inherits the parent's trait~$\beta$ with probability $1\!-\!\mu$. 
With the complementary probability $\mu$, the offspring's payoff sensitivity is slightly mutated by adding some small noise to the parent trait. 
Overall, these assumptions define a dynamical process on the population level, with $\beta$ being the evolving trait. 
This process is straightforward to simulate. 
We give a detailed description of the algorithm in the \methods, and provide the respective code online.

%[ Adaptive dynamics of the prisoner's dilemma and the snowdrift game]
Initially, we assume that the population is monomorphic, with a payoff sensitivity of $\beta\!=\!0$. 
To gain some intuition of the further dynamics, we revisit the previous two examples. 
\textbf{Fig.~\ref{Fig3}A} shows simulations for the prisoner's dilemma. 
Here, we observe that $\beta$ evolves towards ever-increasing values. 
This outcome is consistent with the naive expectation that less erratic learners should have a fitness advantage, and consistent with our analysis in \textbf{Fig.~\ref{Fig2}A--D}. 
For the other example, the snowdrift game, simulations are shown in \textbf{Fig.~\ref{Fig3}C}. 
Also here, larger payoff sensitivities are initially favored. After all, positive $\beta$ values ensure that less time is spent in off-equilibrium states (in fact,  we prove in {\bf SI Section~1B.4} that for non-degenerate payoff matrices, $\beta$ always increases initially). 
On the other hand, the super-game depicted in \textbf{Fig.~\ref{Fig2}H} suggests that at some point, further increases in $\beta$ may no longer be favored, because it leads more sensitive players to spend more time in their less preferred equilibrium.
Indeed, we observe the evolution of payoff sensitivities to stop at some value well below $\beta\!=\!6$.
That is, in contrast to our naive expectation -- but consistent with our analysis in \textbf{Fig.~\ref{Fig2}E--H} -- we observe the evolution of finite $\beta$ values (or, of noisy learning).

%[An intro to PIPs]
We can gain some analytical understanding of these results by considering the respective pairwise invasibility plots (PIPs, \textbf{Fig. \ref{Fig3}B,D}).
These plots depict the direction of evolution as the population size becomes infinitely large, and as mutations become sufficiently rare. 
In that case, we can assume the population to be monomorphic, such that all players have the same (resident) trait of $\beta\!=\!x$. 
To explore whether a mutant with a different trait $y\!\neq\!x$ might invade, we compute the mutant's {\it invasion fitness}, 
\begin{equation}\label{invasionfitness} 
	s_x(y) := \Pi_x(y) - \Pi_x(x).
\end{equation}
In this formula, $\Pi_{x}(y)$ is the payoff of a mutant with payoff sensitivity $\beta\!=\!y$ in a resident population with $\beta\!=\!x$.  
This payoff is given by Eq.~[\ref{averagepayoffs}]. 
The mutant can selectively invade the resident if and only if this invasion fitness is positive. 
In pairwise invasibility plots, the respective region for which $s_x(y)\!>\!0$ is highlighted in color. 
For the prisoner's dilemma, this colored area comprises the entire space above the main diagonal~(\textbf{Fig.~\ref{Fig3}B}). 
This means that independent of the exact payoff sensitivity $x$ of the resident, a mutant with trait $y$ can invade if and only if $y\!>\!x$. 
This implies the evolution of ever-increasing payoff sensitivities. 

In contrast, in the snowdrift game in \textbf{Fig.~\ref{Fig3}D}, mutants with a higher payoff sensitivity are only favored when the resident's sensitivity is sufficiently small (if $x\!<\!x^*$ for some threshold $x^*$). 
We can numerically compute this threshold by setting the so-called \textit{selection gradient} equal to zero (see \methods). 
For the depicted example, we obtain $x^*\!=\!5.58$, in line with the evolutionary simulations. 
The pairwise invasibility plot also indicates that this {\it singular point} is evolutionarily stable. 
Once the population settles at a payoff sensitivity of $\beta\!=\!x^*$, neither a more noisy nor a less noisy learner can invade (as indicated by the white neighborhood around the vertical dashed line).

%[The stag-hunt game]
Interestingly, these two examples of an ever-increasing payoff sensitivity and an evolutionarily stable finite sensitivity are not the only possibilities. 
Instead, \textbf{Fig.~\ref{Fig3}E} shows a third scenario, using a stag-hunt game. 
Here, we observe evolutionary branching.
At first, the monomorphic population again approaches a singular point at  $x^*\!=\!3.01$. 
Once there, both mutant types can invade, those that are more and less noisy learners (indicated by the colored neighborhood of the vertical dashed line in \textbf{Fig.~\ref{Fig3}F}).  
This is the fingerprint of a branching point. 
From here, the population starts diverging in opposite directions. 
There is a sub-population for which sensitivities increase, and another one for which they decrease. 
The subsequent dynamics is more complex to describe analytically; we do so in the  {\bf SI Section~1B.3} and \textbf{Fig.~S2}. 
In the main text, we provide some intuition for why branching occurs in the first place. 
To this end, it is again instructive to look at the stationary distribution of introspection dynamics, for two different sensitivities~$\beta$ (see \textbf{Fig.~S3A--D}). 
When both players use the same $\beta$, a larger~$\beta$ makes players choose the less efficient \textbf{DD} equilibrium more often (because defection is risk-dominant). 
If one player instead has a low $\beta$, both players occasionally coordinate on the more efficient \textbf{CC} equilibrium, while often avoiding harmful mis-coordination on the \textbf{CD}-outcome (see also {\bf SI Section~2C}).

%[Impact on population payoff]
Once the players' sensitivity is allowed to evolve, its dynamics also has important implications on the population's average payoff. 
For example in the prisoner's dilemma, as individuals become more sensitive learners, they increasingly adopt the individually optimal strategy of defection. 
As a result, overall payoffs decrease in time (\textbf{Fig.~S4}, left column). 
For the other two examples, the long-run effects on payoffs are more advantageous. 
Both for the snowdrift and the stag-hunt game, we observe an initially negative but eventually positive trend in the players' payoffs~(\textbf{Fig.~S4}, middle and right column). 
In particular, the evolutionary branching in the stag-hunt game turns out to be both individually and collectively beneficial. 

\begin{figure}[t!]
	\centering
	%[width =  0.9\textwidth]
	\includegraphics[width =  0.48\textwidth]{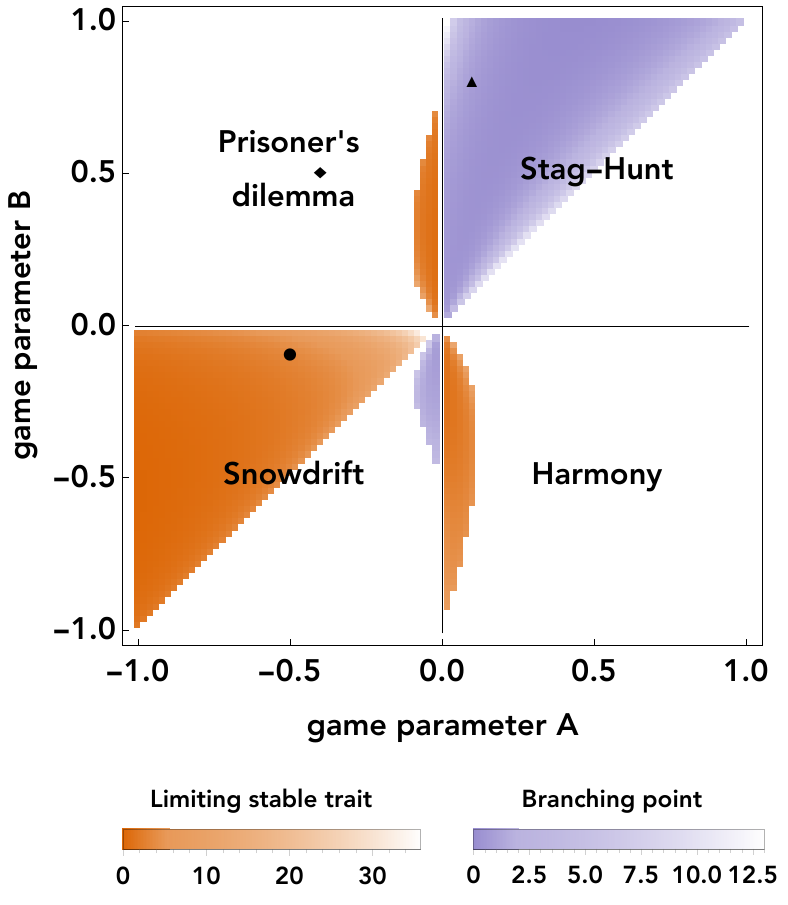}~\\[0.0cm]
	\caption{
		\textbf{Evolutionary dynamics across all social dilemmas.}
		So far, we described three special instances of games; here we systematically repeat this analysis for all $2\!\times\!2$ games. 
		For each pair of game parameters $A$ and $B$, we numerically check whether there exists a finite singular trait value~$\beta^*$.
		If it exists, we explore whether the respective trait value is evolutionarily stable (orange) or a branching point (purple). 
		The color gradient represents the position of the singular trait: the lighter, the larger the value of~$\beta^*$.
		Regions without a finite singular trait are left white. 
		Black symbols indicate the position of the three examples displayed in \textbf{Fig.~\ref{Fig3}}. Note that here we report the lowest singular point (the one which is reached when the process starts at $\beta=0$). However, there are games that permit two singular points. We show those in \textbf{SI Section 2E}.
	}
	\label{Fig4}
\end{figure}

%%[Evolutionary beta for all games]
So far, we have studied the evolution of payoff sensitivity for a few isolated $2\!\times\!2$ games. 
To identify more general patterns, in a next step, we systematically vary the parameters $A$ and $B$ of the stage game. 
For each resulting game, we compute the numerically exact position of the smallest singular point (if such a point exists) and its stability properties.
The result is \textbf{Fig.~\ref{Fig4}}. 
Here, white regions reflect cases without a singular point. 
For the corresponding stage games, we predict sensitivities to increase indefinitely. 
Colored regions correspond to cases with at least one finite singular point. 
Here, we distinguish orange regions (where the smallest point is evolutionarily stable) and purple regions (where we predict branching).
Based on this figure, we can make a number of observations.

%[Prisoner's dilemmas with finite ESS]
First, even though most prisoner's dilemmas lead to ever-increasing payoff sensitivities (as in \textbf{Fig.~\ref{Fig3}A}), there are exceptions. 
These exceptions occur if the absolute value $|A|$ is small (if there is only a small temptation to defect when the co-player cooperates). 
These games permit a finite evolutionarily stable trait value of $\beta$. 
The underlying mechanism is different from the mechanism in the snowdrift game discussed earlier (\textbf{Fig.~\ref{Fig2}E--H}). 
Because $|A|$ is small compared to $B$, a co-player is more likely to occasionally switch to cooperation if the focal player cooperates. 
This positive influence on the co-player's cooperation rate can make it worthwhile to have a finite $\beta$, and to cooperate occasionally~(\textbf{Fig.~S3E--H}). 

Second, in the snowdrift game, we observe evolution towards a finite payoff sensitivity for all games with $B\!>\!A$. 
In addition, as we consider a wider range of parameter values than considered in \textbf{Fig.~\ref{Fig4}}, we find another region exhibiting evolution towards finite $\beta$, see~\textbf{Fig.~S6}. 
This second region is given by $B\!<\!A\!-\!1$; we provide a derivation of these conditions in the \si. 
In both regions, the mechanism favoring finite payoff sensitivities is the same: 
By reducing their sensitivity, learners ensure they spend more time in their individually preferred equilibrium.

%[General dynamics of stag-hunt games]
Third, in the stag-hunt game, we only observe evolutionary branching when it is risk-dominant to defect (that is, for $A\!<\!B$).
If instead cooperation is risk-dominant, the prevalence of \textbf{CC} increases as a function of the players' payoff sensitivity. 
Mutual cooperation is the optimal outcome for both players; hence, an ever-increasing payoff sensitivity is individually and mutually beneficial.

% Harmony games
Finally, in the harmony game quadrant, we observe a small region that permits the evolution of finite payoff sensitivities. 
The existence of this region is particularly puzzling. 
After all, in harmony games, cooperation is both individually and collectively optimal. 
We discuss one of these counterintuitive cases in {\bf Fig.~S3I-L}. 
For this case, the super-game for two given sensitivities can take the form of a prisoner's dilemma: collectively, the two players reach better decisions when they both choose the larger $\beta$ value; yet individually, a player may deviate towards a smaller payoff sensitivity (as this incentivizes the more sensitive player to cooperate even more often). 
Interestingly, however, for each of these harmony games with a finite singular point, there exists in fact a second singular point. Hence, the adaptive dynamics for these harmony games is bistable. If players start out with small payoff sensitivities, they converge to a finite ESS. If their initial sensitivities are sufficiently large, however, their sensitivities further increase indefinitely~({\bf Fig. S5}) -- perhaps the more intuitive outcome in harmony games.

%% MULTIPLAYER GAMES %% 

\subsection*{Beyond pairwise games}
Above, we documented several cases with an unexpected dynamics (e.g., evolution towards finite payoff sensitivities or evolutionary branching).
In each case, the dynamics occurred because at least one player gained some strategic advantage from learning more erratically. 
In general, however, one might expect that these strategic advantages are strongest in pairwise games. 
Among two players, one player's erratic behavior tends to have a comparably strong effect on how the co-player might react. 
This effect could wane in groups of size $N\!>\!2$. 
To further explore the robustness of our results, in the \si{} we study the dynamics of three classes of public goods games. 
The three classes differ in the shape of the public good's benefit function. 
Benefits are either linear, they show diminishing returns~\citep{Hauert:JTB:2006a}, or they take the form of a threshold function~\citep{Archetti:JTB:2011,Pacheco:PRSB:2009}. 
While the computational steps involved in analyzing the adaptive dynamics are the same as before, the mathematical expressions now become more intricate. 
Importantly, however, the qualitative dynamics are similar to before. 
Depending on the payoff function, again we recover the three previously observed cases of indefinite evolution, convergence to a finite evolutionarily stable state, or evolutionary branching~(\textbf{Fig.~S6}, for $N\!=\!3$). 

%% OTHER LEARNING DYNAMICS %% 

\subsection*{Beyond introspection dynamics}
In all our previous examples, we studied the evolution of finite payoff sensitivities under the assumption that individuals learn their strategies based on introspection dynamics. 
However, the learning literature knows of many other learning rules. 
To explore how our results generalize, in the {\bf SI Section~3}, we numerically explore the adaptive dynamics of four other rules. 
All four rules can be regarded as special cases of experience-weighted attraction learning~\cite{Camerer:Econometrica:1999,Dridi:TheoPopBiol:2014,Pangallo:GEB:2022}: stochastic fictitious play, payoff-informed learning, and two different variants of reinforcement learning. 
In each case, we explore the evolution of the player's motivation sensitivity~$\lambda$ (which plays an analogous role to the parameter $\beta$ in introspection dynamics). 

In general, for a given stage game, the four learning rules give rise to pairwise invasibility plots that can be rather different from each other, and different from introspection dynamics~({\bf Figs.~S12-15}). 
However, for each learning rule we find stage games for which $\lambda$ converges towards a finite value. 
Again, these cases are particularly common for stage games with more than one equilibrium (the snowdrift game and the stag-hunt game). 
In those cases, a finite $\lambda$ can again evolve as a means to settle at the player's more preferred equilibrium ({\bf Figs.~S9-S11}). 
In our view, the mere existence of these finite singular points (across all considered learning rules) is remarkable. It suggests a surprisingly pervasive strategic advantage of noisy learning in certain types of stage games.

%%%%%%%%%%%%
%% 	DISCUSSION %%
%%%%%%%%%%%%

\section*{Discussion}

% [Broad overview]
Evolutionary game theory explores how strategies and traits change over time~\cite{Nowak:Science:2004,Broom:book:2013,Traulsen:PTRSB:2023}. 
Traditionally, such an evolutionary approach can be motivated in two ways. 
On the one hand, the respective dynamical equations may describe long-run biological evolution.
Here, individuals with better strategies are assumed to reproduce more often~\cite{Maynard-Smith:Nature:1973,Hofbauer:BC:1982}.
On the other hand, evolutionary models can also be interpreted as a representation of an individual's short-run learning process. 
Such models assume that better strategies are more readily adopted~\cite{Blume:GEB:1993, szabo:PRE:1998, Traulsen:PRE:2006b}. 
Herein, we blend these two interpretations. 
We ask how a long-term biological process may shape the effectiveness of individual learning. 
To this end, we study the evolution of a key quantity in stochastic learning models, the players' payoff sensitivity (which is often referred to as `selection strength' in the related literature). 
This quantity determines to which extent payoff advantages sway individuals to change their strategies during the learning process. 
When individuals are comparably insensitive, they tend to ignore their strategies' payoffs~\citep{Nowak:Nature:2004,Tarnita:JTB:2009,McAvoy:JMB:2015,Allen:Nature:2017,Su:PNAS:2022}.
Here, strategy choices are mostly determined by chance.
As payoff sensitivities become larger, individuals become increasingly swayed by (even minor) payoff differences.

% [Model summary]
Motivated by these observations, herein we interpret payoff sensitivity as an individual trait that is related to the amount of noise in an individual's updating process. 
We use this perspective to study the evolution of noisy learning. 
To this end, we consider a model with two time scales.  
In the short run, individuals have fixed but possibly different payoff sensitivities. 
We explore the resulting learning processes with introspection dynamics~\cite{Couto:NJP:2022, Couto:DGAA:2023,Hauser:Nature:2019,McAvoy:PNASnexus:2022,Wang:PTRSB:2023,huebner:PNAS:2024,Schmid:PlosCB:2022,Ramirez:SciRep:2023}; 
for the $2\!\times\!2$ games we focus on, this process is equivalent to the logit-response dynamics studied in economics~\citep{Couto:DGAA:2023,Alos-Ferrer:GEB:2010}. 
We use this short-term perspective to ask whether individuals who learn less erratically are able to take advantage of their opponents.

% [Results summary]
In the long run, we also allow payoff sensitivities to evolve. 
Here, naively one might expect ever-increasing payoff sensitivities. 
After all, large payoff sensitivities allow individuals to better distinguish profitable strategies from unprofitable ones. 
Surprisingly, however, we identify several important game classes in which payoff sensitivity converges towards a finite value. 
%In some cases, noisy learning seems to act as a commitment device. 
When learning more erratically, individuals occasionally use strategies that run counter to their immediate interests. 
In turn, players who are more sensitive to payoffs are swayed to react and to coordinate on an equilibrium that eventually benefits the noisier learner.
These findings have interesting parallels in the biology literature on the red-king effect~\cite{Bergstrom:PNAS:2003,Damore:Evolution:2011,Veller:pnas:2017}. 
That effect suggests that in mutualisms among two co-evolving species, it can be the slower-evolving species that eventually has an evolutionary advantage.

% prev literature relations
Our model is also directly related to two interesting studies by Szab\'o, Vukov, and Szolnoki~\cite{Szabo:EPL:2009, Szolnoki:PRE:2009c}. 
They also consider a model in which two traits co-evolve. 
These traits are the individuals' strategies and a noise parameter that is inversely related to our payoff sensitivity. 
Our work deviates from these studies in many ways. 
For example, we look at well-mixed populations while they study behavior on a regular lattice; we treat payoff sensitivity as a continuous variable while they consider a discrete set of possible noise parameters; we explore the entire range of $2\!\times\!2$ games, whereas they consider a one-dimensional subspace of games. 
Most importantly, however, they consider simulations in which strategies and the noise parameter evolve at the same rate. 
Instead, we assume a separation of timescales. 
As a result, they do not address most of the questions that are central to our approach. 
For example, because they do not allow for lasting differences in an individual's learning capabilities, they do not explore to which extent more sensitive learners are able to outperform their opponents. 
This question plays a crucial role in our setup, and it drives all of our subsequent results (such as the emergence of evolutionary branching, which they do not observe).

%mixed games: Extension of the considered game sets
We have already sketched possible generalizations of our baseline model. 
In particular, in the \si{} we provide more details on how the model can be extended to cover multiplayer interactions~({\bf SI Section~2F}) or alternative learning rules~({\bf SI Section~3}).  
Several other extensions are possible too.
For example, herein, we considered individuals who engage in a series of one-shot (non-repeated) games. 
In {\bf SI Section~2G}, we sketch instead how to capture strategies of direct reciprocity~\citep{Garcia:FRAI:2018} within our setup. To this end, we consider the pairwise dynamics when individuals choose between ALLD and Tit-for-Tat.
Surprisingly, we find that depending on how individuals discount the future, already these two strategies can generate all of our previously described qualitative scenarios (see \textbf{Fig.~S7}).
Future work could provide a more comprehensive analysis that also includes other repeated-game strategies. 
Other more complex setups that one could study within our framework are asymmetric games~\citep{Hauser:Nature:2019,Wang:PTRSB:2023,huebner:PNAS:2024}, and heterogeneous environments \cite{Balliet:PSPR:2017, Colnaghi:PNAS:2023}.
Such an approach could help us understand how people have developed psychological mechanisms to cope with the variability of situations they face.

% structured pop
Yet another interesting direction would be to study the evolution of noisy learning in structured rather than well-mixed populations. 
A framework of adaptive dynamics in such populations is readily available \cite{Allen:AmNat:2013, Hauert:PNAS:2021}.
The literature shows that network structure strongly affects the way individuals cooperate \cite{Ohtsuki:Nature:2006, Santos:Nature:2008,Perc:JRSI:2013}. 
With a framework similar to ours, one could explore how network structure affects the way individuals learn. 
This work could investigate, for example, whether heterogeneous networks promote heterogeneity in people's learning mechanisms. 
Finally, our results can also find future applications in multi-agent reinforcement learning. 
Specifically, one could conceive extensions where agents dynamically adapt their learning rates as a function of the dilemmas faced \cite{Bowling:AI:2002}.

\section*{Materials and Methods}

We provide a detailed account of both the short-run learning dynamics and the long-run evolutionary dynamics in the \si. 
In the following, we briefly sketch our methods for the long-run evolutionary dynamics. 
For that, we use a combination of individual-based simulations and a theoretical analysis in the infinite population limit. 

\subsection*{Evolutionary simulations}
Our evolutionary simulations use a similar setup as D\"obeli et al~\cite{Doebeli:Science:2004}.
We start with a monomorphic population with trait $\beta=0$. 
At each time step, two individuals, $i$ and $j$, are randomly chosen from the population. 
These two individuals then compete for reproduction. 
To model this competition, each of the two individuals independently engages in a sample of three encounters with other random individuals.
For each encounter, their payoff is computed with \eqref{averagepayoffs}. 
By averaging over the three realizations, we compute each player's average payoff $\pi_i$ and $\pi_j$.
With probability $1/\big(1\!+\!\exp[-30 (\pi_i\!-\!\pi_j)]\big)$, individual $i$ then reproduces and $j$ dies. 
Otherwise, individual $j$ reproduces and $i$ dies.
During reproduction, there is a small chance that an offspring inherits an imperfect copy of the parent trait (with a mutation rate of $\mu\! =\! 0.005$).
In that case, the mutation is uniformly distributed around the parent trait in an interval of size 0.1.
We iterate this simulation step a large number of times.
At regular time intervals, we record the composition of the population and the average cooperation rates and payoffs.  
For the simulations in \textbf{Fig.~\ref{Fig3}}, we consider comparably large populations ($10^4$ in the prisoner's dilemma and the snowdrift game; $2\!\cdot\!10^4$ in the stag-hunt game). 

\subsection*{Theoretical analysis}
For our analytical results, we use the adaptive dynamics formalism~\cite{Geritz:PRL:1997, geritz:EER:1998, Hofbauer:book:1998}. 
Here, we assume infinitely large populations and rare mutations of infinitesimal magnitude. 
Under these assumptions, we can focus on mutant traits close to the resident, such that $y\!\approx\!x$. 
A key quantity to study such mutants is the \textit{local fitness gradient} or \textit{selection gradient} $D(x)$, defined by
\begin{equation}\label{ch3_grad} 
	D(x) := \left[ \frac{\partial s_x(y)}{\partial y } \right]_{y=x}.
\end{equation}
The selection gradient determines the direction of evolutionary change. 
When $D(x)$ is positive (negative), mutants with a slightly higher (lower) trait value than $x$ replace the resident population.
In that case, the resident trait value $x$ grows (decreases).
A point $x^*$ that satisfies $D(x^*)=0$ is called a \textit{singular point}. 
A singular point is called \textit{convergence stable} if 
\begin{equation}
	\left[\frac{d D(x)}{dx}\right]_{x=x^*} \!<\!0.
\end{equation}
This condition determines whether the singular point can be reached at all, starting from a resident population sufficiently close by. 
Moreover, the singular point is \textit{evolutionary stable} if
\begin{equation}
	\left[\frac{\partial ^2 s_x(y)}{\partial y^2}\right]_{x=y=x^*}\!<\!0.
\end{equation} 
In that case, the fitness function has a local maximum at $x^*$, and the singular point corresponds to an evolutionary endpoint.
If the singular point is at a fitness minimum, however, this may indicate the presence of an \textit{evolutionary branching point}. 
For \textbf{Fig.~\ref{Fig4}}, we verify the respective conditions numerically.\\

\noindent
{\bf Data, Materials, and Software Availability.} Simulations have been run with MATLAB. 
For our analytical results, we used Mathematica. 
The respective simulation data and source code can be found in this \href{https://osf.io/ukcnp/?view_only=c37ac22cb83c4c8a9082c9c9859466c1}{OSF repository}.\\

\noindent
{\bf Author contributions.} M.C.C., F.P.S. and C.H. designed research; M.C.C. performed research and analyzed data; and M.C.C., F.P.S. and C.H. wrote the paper.\\

\noindent
{\bf Author declaration.} The authors declare no competing interest.\\

\noindent
{\bf Acknowledgments.} M.C.C and C.H. acknowledge generous support by the European Research Council Starting Grant 850529: E-DIRECT.
F.P.S acknowledges funding from the Dutch Research Council (NWO) through the project with file number OCENW.M.22.322 of the research programme Open Competitie ENW.

\newpage

% Bibliography
\bibliographystyle{unsrt}
\bibliography{bibtex}

%\end{multicols}

%\includepdf[pages=-]{SI.pdf}

\end{document}